\documentclass[12pt]{article}
\usepackage{amsmath,amsthm,amscd,amssymb}
\usepackage{tikz}
\usepackage{graphicx,psfrag,epsf}
\usepackage{enumerate}
\usepackage{natbib}

\usepackage{psfig}
\usepackage{epsfig}
\usepackage{epstopdf}

\usepackage{hyperref}

\usepackage{url}
\usepackage{pdfsync}
\usepackage{blkarray}
\def\Big#1{\makebox(-1,1){\Large#1}}

\usepackage{algorithmic}
\usepackage{algorithm}

\usepackage{colortbl}

\renewcommand{\epsilon}{\varepsilon}
\renewcommand{\Sigma}{\varSigma}

\title{Split Hamiltonian Monte Carlo}

\author{Babak Shahbaba\footnote{Department of Statistics, University of California, Irvine, USA.} , Shiwei Lan$^{*}$, Wesley O. Johnson$^{*}$, and Radford M. Neal\footnote{Department of Statistics and Department of Computer Science, University of Toronto, Canada.}}

\date{\today}

\begin{document}

\def\spacingset#1{\renewcommand{\baselinestretch}%
{#1}\small\normalsize} \spacingset{1}

\maketitle

\bigskip
\begin{abstract}\noindent
We show how the Hamiltonian Monte Carlo algorithm can sometimes be
speeded up by ``splitting'' the Hamiltonian in a way that allows much
of the movement around the state space to be done at low computational
cost.  One context where this is possible is when the log density of
the distribution of interest (the potential energy function) can be
written as the log of a Gaussian density, which is a quadratic
function, plus a slowly varying function.  Hamiltonian dynamics for
quadratic energy functions can be analytically solved.  With the
splitting technique, only the slowly-varying part of the energy needs
to be handled numerically, and this can be done with a larger stepsize
(and hence fewer steps) than would be necessary with a direct
simulation of the dynamics.  Another context where splitting helps is
when the most important terms of the potential energy function and its gradient can be
evaluated quickly, with only a slowly-varying part requiring costly
computations.  With splitting, the quick portion can be handled with a
small stepsize, while the costly portion uses a larger stepsize.  We
show that both of these splitting approaches can reduce the
computational cost of sampling from the posterior distribution for a
logistic regression model, using either a Gaussian approximation centered on the posterior mode, or a Hamiltonian split into a term that depends on only a small number of critical cases, and another term
that involves the larger number of cases whose influence on
the posterior distribution is small. Supplemental materials for this paper are 
available online. 

\vspace{8pt}


\end{abstract}

\spacingset{1.45}
\section{Introduction}
\label{sec:intro}
The simple Metropolis algorithm \citep{metropolis59} is often
effective at exploring low-dimensional distributions, but it can be
very inefficient for complex, high-dimensional distributions ---
successive states may exhibit high autocorrelation, due to the random
walk nature of the movement.  Faster exploration can be obtained using
Hamiltonian Monte Carlo, which was first introduced by \cite{duane87},
who called it ``hybrid Monte Carlo'', and which has been recently
reviewed by \cite{neal10}.  Hamiltonian Monte Carlo (HMC) reduces the
random walk behavior of Metropolis by proposing states that are
distant from the current state, but nevertheless have a high
probability of acceptance.  These distant proposals are found by
numerically simulating Hamiltonian dynamics for some specified amount
of fictitious time.

For this simulation to be reasonably accurate (as required for a high
acceptance probability), the stepsize used must be suitably small.
This stepsize determines the number of steps needed to produce the
proposed new state.  Since each step of this simulation requires a
costly evaluation of the gradient of the log density, the stepsize is the main determinant of computational cost.

In this paper, we show how the technique of ``splitting'' the
Hamiltonian \citep{leimkuhler04, neal10} can be used to reduce the
computational cost of producing proposals for Hamiltonian Monte Carlo.
In our approach, splitting separates the Hamiltonian, and consequently the simulation
of the dynamics, into two parts.  We discuss two contexts in which one
of these parts can capture most of the rapid variation in the energy
function, but is computationally cheap.  Simulating the other,
slowly-varying, part requires costly steps, but can use a large
stepsize. The result is that fewer costly gradient evaluations are
needed to produce a distant proposal. We illustrate these splitting methods using logistic regression models. Computer programs for our methods are publicly available from \url{http://www.ics.uci.edu/~babaks/Site/Codes.html}.

Before discussing the splitting technique, we provide a brief overview
of HMC. \citep[See][for an extended review of HMC.]{neal10} To begin,
we briefly discuss a physical interpretation of Hamiltonian
dynamics. Consider a frictionless puck that slides on a surface of
varying height. The state space of this dynamical system consists of its
\emph{position}, denoted by the vector $q$, and its momentum (mass,
$m$, times velocity, $v$), denoted by a vector $p$. Based on $q$ and
$p$, we define the \emph{potential energy}, $U(q)$, and the
\emph{kinetic energy}, $K(p)$, of the puck. $U(q)$ is proportional to
the height of the surface at position $q$. The kinetic energy is
$m|v|^2/2$, so $K(p) = |p|^{2}/(2m)$. As the puck moves on an upward
slope, its potential energy increases while its kinetic energy
decreases, until it becomes zero. At that point, the puck slides back down,
with its potential energy decreasing and its kinetic energy
increasing.

The above dynamic system can be represented by a function of $q$ and $p$
known as the \emph{Hamiltonian}, which for HMC is usually defined as
the sum of a potential energy, $U$, depending only on the position and a
kinetic energy, $K$, depending only on the momentum:
\begin{eqnarray}\label{hamiltonian}
H(q, p) & = & U(q) + K(p)
\end{eqnarray}
The partial derivatives of $H(q, p)$ determine how $q$ and $p$ change over 
time, according to \emph{Hamilton's equations}:\vspace{-6pt}
\begin{eqnarray}\begin{array}{ccrcr}
\displaystyle \frac{dq_{j}}{dt} 
               & = & \displaystyle \frac{\partial H}{\partial p_{j}} 
               & = & \displaystyle \frac{\partial K}{\partial p_{j}} \\[16pt]
\displaystyle \frac{dp_{j}}{dt} 
               & = & \displaystyle - \frac{\partial H}{\partial q_{j}} 
               & = & \displaystyle - \frac{\partial U}{\partial q_{j}} 
\end{array}\label{HamiltonsEquations}\end{eqnarray}
These equations define a mapping $T_{s}$ from the state at some time $t$ 
to the state at time $t+s$. 

We can use Hamiltonian dynamics to sample from some distribution of
interest by defining the potential energy function to be minus the log of
the density function of this distribution (plus any constant).  The
position variables, $q$, then correspond to the variables of interest.  
We also introduce fictitious momentum variables, $p$, of the same
dimension as $q$, which will have a distribution defined by the 
kinetic energy function.  The joint density of $q$ and $p$ is
defined by the Hamiltonian function as
\begin{eqnarray*}
P(q, p) & = & \frac{1}{Z}\,\exp \Big[-H(q, p) \ \Big]
\end{eqnarray*}
When $H(q, p) \,=\, U(q)+K(p)$, as we assume in this paper, we have
\begin{eqnarray*}
P(q, p) & = & \frac{1}{Z}\,\exp \Big[\ U(q) \ \Big] \,\exp \ \Big[ -K(p) \ \Big]
\end{eqnarray*}
so $q$ and $p$ are independent.  Typically, $K(p) \,=\,
p^{T}M^{-1}p\,/\,2$, with $M$ usually being a diagonal matrix with
elements $m_{1}, \ldots, m_{d}$, so that $K(p) \,=\, \sum_{i}
p^{2}_{i}/2m_{i}$.  The $p_j$ are then independent and Gaussian with
mean zero, with $p_j$ having variance $m_j$.

In applications to Bayesian statistics, $q$ consists of the model parameters
(and perhaps latent variables), and 
our objective is to sample from the posterior distribution for $q$ given
the observed data $D$. To this end, we set
\begin{eqnarray*}
U(q) & = & -\log [P(q)L(q|D)]
\end{eqnarray*}
where $P(q)$ is our prior and $L(q|D)$ is the likelihood function given data 
$D$. 

Having defined a Hamiltonian function corresponding to the
distribution of interest (e.g., a posterior distribution of model
parameters), we could in theory use Hamilton's equations, applied for
some specified time period, to propose a new state in the Metropolis
algorithm.  Since Hamiltonian dynamics leaves invariant the value of
$H$ (and hence the probability density), and preserves volume, this
proposal would always be accepted.  (For a more detailed explanation,
see \cite{neal10}.)

In practice,
however, solving Hamiltonian's equations exactly is too hard, so we
need to approximate these equations by discretizing time, using some
small step size $\epsilon$. For this purpose, the \emph{leapfrog}
method is commonly used.  It consists of iterating the following
steps:
\begin{eqnarray}\label{leapfrog}
p_{j}(t+\epsilon/2) & = &  p_{j}(t) \ -\
   (\epsilon/2) \frac{\partial U}{\partial q_{j}}(q(t))\nonumber\\[3pt]
q_{j}(t+\epsilon) & = &  q_{j}(t) \ +\
   \epsilon \frac{\partial K}{\partial p_{j}}(p(t+\epsilon/2))\\[3pt]
p_{j}(t+\epsilon) & = &  p_{j}(t+\epsilon/2) \ -\
   (\epsilon/2) \frac{\partial U}{\partial q_{j}}(q(t+\epsilon))\nonumber
\end{eqnarray}
In a typical case when $K(p) \,=\, \sum_{i} p^{2}_{i}/2m_{i}$, the
time derivative of $q_j$ is $\partial K/\partial p_{j} \,=\,p_{j}/m_{j}$.
The computational cost of a leapfrog step will then usually be 
dominated by evaluation of $\partial U/\partial q_j$.

We can use some number, $L$, of these leapfrog steps, with some
stepsize, $\epsilon$, to propose a new state in the Metropolis
algorithm.  We apply these steps starting at the current state
$(q,p)$, with fictitious time set to $t=0$.  The final state, at time
$t=L\epsilon$, is taken as the proposal, $(q^*,p^*)$.  (To make the
proposal symmetric, we would need to negate the momentum at the end of
the trajectory, but this can be omitted when, as here, $K(p) = K(-p)$
and $p$ will be replaced (see below) before the next update.)  This
proposal is then either accepted or rejected (with the state remaining 
unchanged), with the acceptance probability being 
\begin{eqnarray*}
   \min [1,\, \exp(-H(q^{*}, p^{*})+H(q, p))]
   & = & \min [1,\, \exp(-U(q^{*})+ U(q) - K(p^{*})+K(p))]
\end{eqnarray*}

These Metropolis updates will leave $H$ approximately constant, and
therefore do not explore the whole joint distribution of $q$ and $p$.
The HMC method therefore alternates these Metropolis updates 
with updates in which the momentum is sampled from its distribution (which
is independent of $q$ when $H$ has the form in Eq.~(\ref{hamiltonian})).  
When $K(p) \,=\, \sum_{i}
p^{2}_{i}/2m_{i}$, each $p_{j}$ is sampled independently from the Gaussian
distribution with mean zero and variance $m_j$.

\begin{figure}[t]
\begin{center}
\centerline{
  \includegraphics[width=3in]{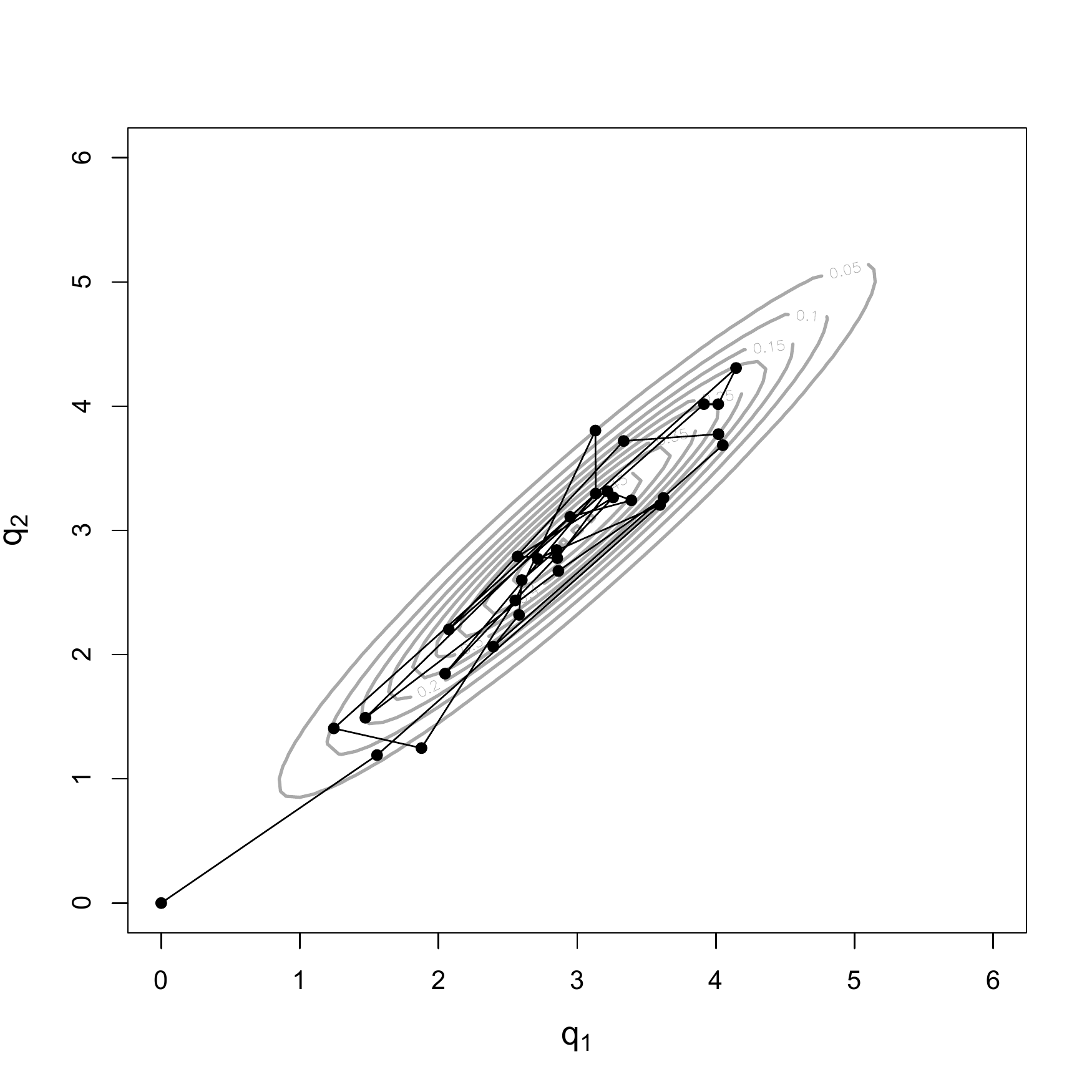}
  \includegraphics[width=3in]{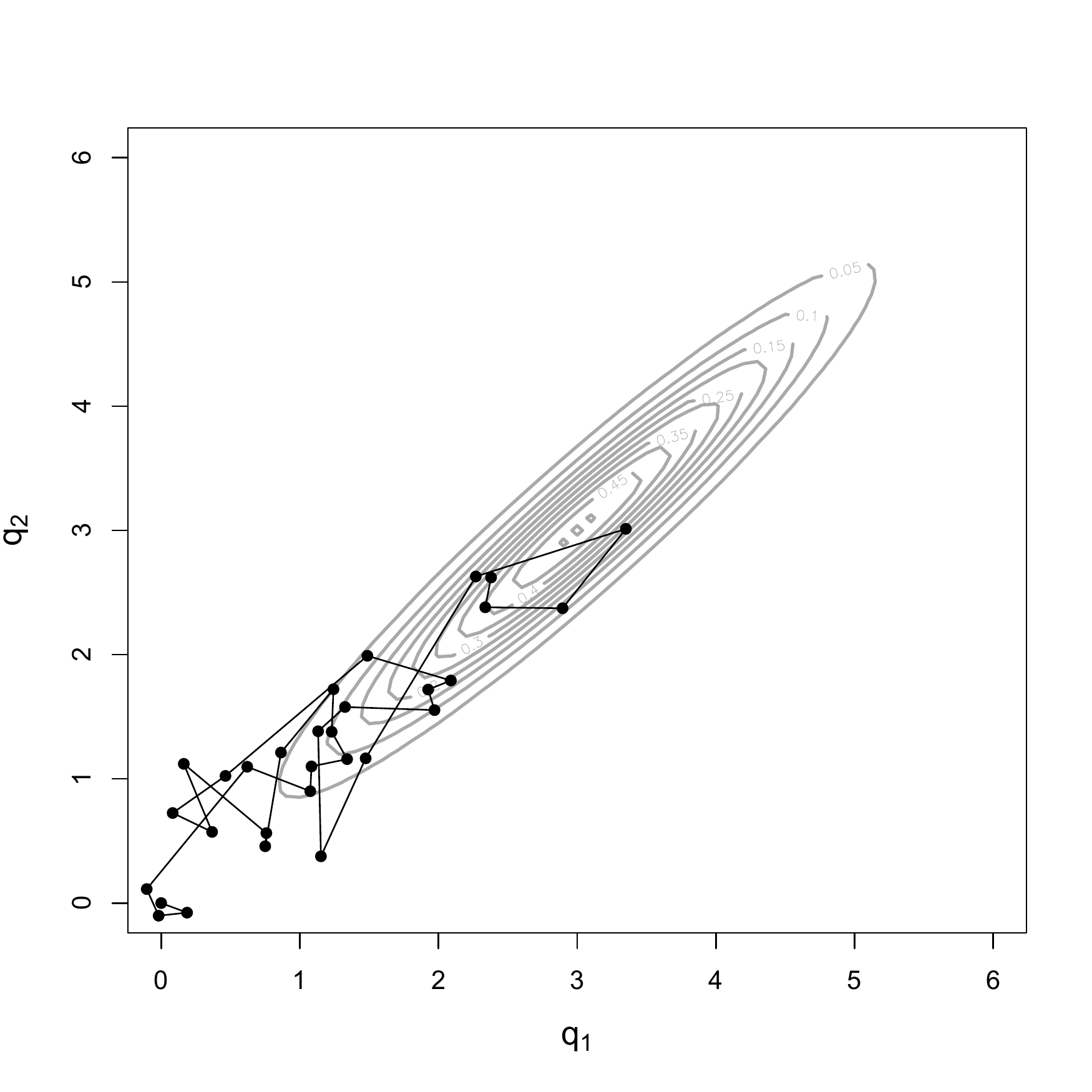}}

\caption{Comparison of Hamiltonian Monte Carlo (HMC) and Random Walk
Metropolis (RWM) when applied to a bivariate
normal distribution. Left plot: The first 30
iterations of HMC with 20 leapfrog steps.
Right plot: The first 30 iterations of RWM
with 20 updates per iterations.}
\label{sim1Plot}
\end{center}
\end{figure}

As an illustration, consider sampling from the following bivariate normal 
distribution
\begin{eqnarray*}
   q & \sim & N(\mu, \Sigma), \quad 
   \textrm{with } \mu = \binom{3}{3}\ \textrm{and}\
                  \Sigma = \binom{1 \quad 0.95}{0.95\quad 1}
\end{eqnarray*}
For HMC, we set $L=20$ and $\epsilon=0.15$. The left plot in
Figure~\ref{sim1Plot} shows the first 30 states from an HMC run started
with $q=(0,0)$. The 
density contours of the bivariate normal distribution are shown as gray
ellipses. The right plot shows every $20^{th}$ state from the first
600 iterations of a run of a simple random walk Metropolis (RWM) algorithm.
(This takes time comparable to that for the HMC run.) The
proposal distribution for RWM is a bivariate normal with the current
state as the mean, and $0.15^{2}I_{2}$ as the covariance matrix. (The
standard deviation of this proposal is the same as the stepsize of
HMC.) Figure \ref{sim1Plot} shows that HMC explores the distribution more efficiently, with
successive samples being further from each other, and autocorrelations
being smaller. For an extended review of HMC, its properties, and
its advantages over the simple random walk Metropolis algorithm, see
\cite{neal10}.

In this example, we have assumed that one leapfrog step for HMC (which
requires evaluating the gradient of the log density) takes
approximately the same computation time as one Metropolis update
(which requires evaluating the log density), and that both move
approximately the same distance.  The benefit of HMC comes from this
movement being systematic, rather than in a random walk.\footnote{Indeed,
in this two-dimensional example, it is better to use Metropolis with
a large proposal standard deviation, even though this leads to a
low acceptance probability, because this also avoids a random walk.
However, in higher-dimensional problems with more than one highly-confining
direction, a large proposal standard deviation leads to such a
low acceptance probability that this strategy is not viable.}
We now
propose a new approach called Split Hamiltonian Monte Carlo (Split
HMC), which further improves the performance of HMC by modifying how
steps are done, with the effect of reducing the time for one step or
increasing the distance that one step moves.

\section{Splitting the Hamiltonian}

As discussed by \cite{neal10}, variations on HMC can be obtained by
using discretizations of Hamiltonian dynamics derived by ``splitting'' 
the Hamiltonian, $H$, into several terms:
\begin{eqnarray*}
H(q, p) & = & H_{1}(q, p) \,+\, H_{2}(q, p) \,+\, \cdots\, +\, H_{K}(q, p)
\end{eqnarray*}
We use $T_{i, t}$, for $i=1, \ldots, k$ to denote the mapping defined
by $H_{i}$ for time $t$. Assuming that we can implement Hamiltonian
dynamics $H_{k}$ exactly, the composition $T_{1, \epsilon} \circ T_{2,
\epsilon} \circ \ldots \circ T_{k, \epsilon}$ is a valid
discretization of Hamiltonian dynamics based on $H$ if the $H_{i}$ are
twice differentiable \citep{leimkuhler04}. This discretization is
symplectic and hence preserves volume. It will also be reversible if
the sequence of $H_{i}$ are symmetric: $H_{i}(q, p) = H_{K - i+1}(q,
p)$.

Indeed, the leapfrog method (\ref{leapfrog}) can be regarded as a
symmetric splitting of the Hamiltonian $H(q, p) = U(q) + K(p)$ as
\begin{eqnarray}\label{split1}
H(q, p) & = & U(q)/2 \,+\, K(p) \,+\, U(q)/2
\end{eqnarray}
In this case, $H_{1}(q, p) = H_{3}(q, p) = U(q)/2$ and $H_{2}(q, p) =
K(p)$. Hamiltonian dynamics for $H_{1}$ is
\begin{eqnarray*}
\frac{dq_{j}}{dt} & = & \frac{\partial H_{1}}{\partial p_{j}} \ \ =\ \ 0 \\[4pt]
\frac{dp_{j}}{dt} & = & - \frac{\partial H_{1}}{\partial q_{j}} 
                  \ \ =\ \, -\frac{1}{2}\, \frac{\partial U}{\partial q_{j}}
\end{eqnarray*}
which for a duration of $\epsilon$ gives the first part of a leapfrog step.
For $H_{2}$, the dynamics is
\begin{eqnarray*}
\frac{dq_{j}}{dt} & = & \frac{\partial H_{2}}{\partial p_{j}} 
                  \ \ =\ \ \frac{\partial K}{\partial p_{j}} \\[4pt]
\frac{dp_{j}}{dt} & = & - \frac{\partial H_2}{\partial q_{j}} \ \ =\ \ 0
\end{eqnarray*}
For time $\epsilon$, this gives the second part of the leapfrog step. 
Hamiltonian dynamics for $H_{3}$ is the same as that for $H_{1}$ since
$H_{1} = H_{3}$, giving the the third part of the leapfrog step.

\subsection{Splitting the Hamiltonian when a partial analytic solution \\[-3pt]
            is available}
  
Suppose the potential energy $U(q)$ can be written as $U_{0}(q) +
U_{1}(q)$. Then, we can split $H$ as \begin{eqnarray}\label{split2}
H(q, p) & = & U_{1}(q)/2 + [U_{0}(q) + K(p)] + U_{1}(q)/2
\end{eqnarray} Here, $H_{1}(q, p) = H_{3}(q, p) = U_{1}(q)/2$ and
$H_{2}(q, p) = U_{0}(p) + K(p)$. The first and the last terms in this
splitting are similar to Eq.~(\ref{split1}), except that $U_{1}(q)$
replaces $U(q)$, so the first and the last part of a leapfrog step
remain as before, except that we use $U_{1}(q)$ rather than $U(q)$ to
update $p$. Now suppose that the middle part of the leapfrog, which is
based on the Hamiltonian $U_{0}(q) + K(p)$, can be handled
analytically --- that is, we can compute the exact dynamics for any
duration of time.  We hope that since this part of the simulation
introduces no error, we will be able to use a larger step size, and
hence take fewer steps, reducing the computation time for the
dynamical simulations.

We are mainly interested in situations where $U_{0}(q)$ provides a
reasonable approximation to $U(q)$, and in particular on Bayesian
applications, where we approximate $U$ by focusing on the the
posterior mode, $\hat{q}$, and the second derivatives of $U$ at that
point.  We can obtain $\hat{q}$ using fast methods such as 
Newton-Raphson iteration when analytical solutions are not available. We then
approximate $U(q)$ with $U_{0}(q)$, the energy function for
$N(\hat{q}, \mathcal{J}^{-1}(\hat{q}))$, where $\mathcal{J}(\hat{q})$
is the Hessian matrix of $U$ at $\hat{q}$.  Finally, we set $U_{1}(q)
= U(q) - U_{0}(q)$, the error in this approximation.

\cite{beskos11} have recently proposed a similar splitting strategy
for HMC, in which a Gaussian component is handled analytically, in the
context of high-dimensional approximations to a distribution on an
infinite-dimensional Hilbert space.  In such applications, the
Gaussian distribution will typically be derived from the problem
specification, rather than being found as a numerical approximation,
as we do here.

Using a normal approximation in which $U_0(q) = \frac{1}{2} (q -
\hat{q})^{T} \mathcal{J} (\hat{q}) (q - \hat{q})$, and letting $K(p) =
\frac{1}{2} p^Tp$ (the energy for the standard normal distribution),
$H_{2}(q, p) = U_0(q) + K(p)$ in Eq.~(\eqref{split2}) will be
quadratic, and Hamilton's equations will be a system of first-order
linear differential equations that can be handled analytically
\citep{polyanin02}. Specifically, setting $q^{*} = q - \hat{q}$, 
the dynamical equations can be written as follows:
\begin{eqnarray*}\label{HDU0}
\frac {d}{dt}\begin{bmatrix}
q^{*}(t)\\
p(t)
\end{bmatrix}
& = & \begin{bmatrix}
0 & I\\
- \mathcal{J} (\hat{q}) & 0
\end{bmatrix}
\begin{bmatrix}
q^{*}(t)\\
p(t)
\end{bmatrix}
\end{eqnarray*}
where $I$ is the identity matrix. Defining $X\,=\,(q,\,p)$, this can be written
as $\displaystyle \frac{d}{d t} X(t) = A X(t)$, where
\begin{eqnarray*}
A & = & \begin{bmatrix}
  0 & I\\
  - \mathcal{J} (\hat{q}) & 0
\end{bmatrix}
\end{eqnarray*}
The solution of this system is $X(t)\, =\, e^{At}\,X_0$, where $X_0$ is the
initial value at time $t=0$, and $e^{At} = I + (At) + (At)^2/2! + \cdots$ is
a matrix exponential.  This can be simplified by diagonalizing the 
coefficient matrix $A$ as
\begin{eqnarray*}
A & = & \Gamma D \Gamma^{-1}
\end{eqnarray*}
where $\Gamma$ is invertible and $D$ is a diagonal matrix.  The system of
equations can then be written as 
\begin{eqnarray*}
\frac{d}{d t} X(t) & = & \Gamma D \Gamma^{-1} X(t)
\end{eqnarray*}
Now, let $Y(t) = \Gamma^{-1} X(t)$. Then, 
\begin{eqnarray*}
\frac{d}{d t} Y(t) & = & D Y(t)
\end{eqnarray*}
The solution for the above equation is $Y(t) = e^{Dt}\, Y_0$, where 
$Y_{0} = \Gamma^{-1} X_{0}$. Therefore, 
\begin{eqnarray*}
X(t) & = & \Gamma e^{Dt}\, \Gamma^{-1}\, X_0
\end{eqnarray*}
and $e^{Dt}$ can be easily computed by simply exponentiating the diagonal
elements of $D$ times $t$.

The above analytical solution is of course for the middle part
(denoted as $H_{2}$) of Eq.~\eqref{split2} only. We still need to
approximate the overall Hamiltonian dynamics based on $H$, using the
leapfrog method. Algorithm~1 shows the corresponding
leapfrog steps --- after an initial step of size
$\epsilon/2$ based on $U_{1}(q)$, we obtain the exact solution for a time
step of $\epsilon$ based on $H_{2}(q, p) = U_0(q) + K(p)$, and finish 
by taking another step of size $\epsilon/2$ based on $U_{1}(q)$.

\begin{figure}
\begin{minipage}[t]{0.48\linewidth}
\textbf{Algorithm 1:}\ \ Leapfrog for split Hamiltonian Monte Carlo with a 
         partial analytic solution.\\[-6pt]
%\label{algAnalytical}
\rule{\linewidth}{1pt}\vspace{-28pt}
\begin{algorithmic}\vspace{4pt}
\STATE $R \leftarrow \Gamma e^{D\epsilon} \Gamma^{-1}$
\STATE Sample initial values for $p$ from $N(0,I)$
\FOR{$\ell = 1$ to $L$ }
\STATE $p \leftarrow p - 
         (\epsilon/2) \displaystyle \frac{\partial U_{1}}{\partial q}$
\STATE $q^{*} \leftarrow q - \hat{q}$
\STATE $X_{0} \leftarrow (q^{*}, p)$
\STATE $(q^{*}, p) \leftarrow R X_{0}$ 
\STATE $q \leftarrow q^{*} + \hat{q}$ 
\STATE $p \leftarrow p -
          (\epsilon/2) \displaystyle \frac{\partial U_{1}}{\partial q}$
\ENDFOR
\end{algorithmic}
\end{minipage}
\vspace{0.01\linewidth}
\rule[-3.43in]{1pt}{3.55in}
\vspace{0.02\linewidth}
\begin{minipage}[t]{0.49\linewidth}
\textbf{Algorithm 2:}\ \ Nested leapfrog for split Hamiltonian Monte Carlo with
         splitting of data.\\[-6pt]
%\label{algNestedLeaps}
\rule{\linewidth}{1pt}\vspace{-28pt}
\begin{algorithmic}\vspace{4pt}
\STATE Sample initial values for $p$ from $N(0,I)$
\FOR{$\ell = 1$ to $L$ }
\STATE $p \leftarrow p -  
          (\epsilon/2) \displaystyle \frac{\partial U_{1}}{\partial q}$
\FOR{$m = 1$ to $M$ } 
\STATE $p \leftarrow p -  
          (\epsilon/2M) \displaystyle \frac{\partial U_{0}}{\partial q}$
\STATE $q \leftarrow q + (\epsilon/M) p$
\STATE $p \leftarrow p -  
          (\epsilon/2M) \displaystyle \frac{\partial U_{0}}{\partial q}$
\ENDFOR
\STATE $p \leftarrow p - 
          (\epsilon/2) \displaystyle \frac{\partial U_{1}}{\partial q}$
\ENDFOR
\end{algorithmic}
\end{minipage}
\rule{\linewidth}{1pt}\vspace{-8pt}
\end{figure}

\subsection{Splitting the Hamiltonian by splitting the data}

The method discussed in the previous section requires that we be able
to handle the Hamiltonian $H_{2}(q, p) = U_{0}(q) + K(p)$
analytically. If this is not so, splitting the Hamiltonian in this way
may still be beneficial if the computational cost for $U_{0}(q)$ is
substantially lower than for $U(q)$. In these situations, we can use
the following split: 
\begin{eqnarray}\label{split3} 
H(q, p) & = & 
  U_{1}(q)/2 + \sum_{m=1}^{M}[U_{0}(q)/2M + K(p)/M + U_{0}(q)/2M] + U_{1}(q)/2
\end{eqnarray} 
for some $M > 1$. The above discretization can be considered as a
nested leapfrog, where the outer part takes half steps to update $p$
based on $U_{1}$ alone, and the inner part involves $M$ leapfrog steps
of size $\epsilon/M$ based on $U_{0}$.  Algorithm~2
implements this nested leapfrog method.

For example, suppose our statistical analysis involves a large data
set with many observations, but we believe that a small subset of data
is sufficient to build a model that performs reasonably well (i.e.,
compared to the model that uses all the observations). In this case,
we can construct $U_{0}(q)$ based on a small part of the observed
data, and use the remaining observations to construct $U_{1}(q)$. 
If this strategy is successful, we will able to use a large stepsize
for steps based on $U_1$, reducing the cost of a trajectory computation.

In detail, we divide the observed data, $y$, into two subsets: $R_{0}$, which
is used to construct $U_{0}(q)$, and $R_{1}$, which is used to
construct $U_{1}$:
\begin{eqnarray} \label{eq:splitData}
U(\theta) & = &  U_{0}(\theta) + U_{1}(\theta) \nonumber \\
U_{0}(\theta) & = & -\log[P(\theta)] - \sum_{i \in R_{0}} \log[P(y_{i}| \theta)]\\
U_{1}(\theta) & = & - \sum_{i' \in R_{1}} \log[P(y_{i'}| \theta)] \nonumber
\end{eqnarray}
Note that the prior appears in $U_{0}(\theta)$ only. 

\cite{neal10} discusses a related strategy for splitting the
Hamiltonian by splitting the observed data into multiple
subsets. However, instead of randomly splitting data, as proposed
there, here we split data by building an initial model based on the
maximum a posterior (MAP) estimate, $\hat{q}$, and use this model to
identify a small subset of data that captures most of the information
in the full data set.  We next illustrate this approach for logistic
regression models.

%%%%%%%%%%%%%%%%%%%%%%%%%%%%%%

\section{Application of Split HMC to logistic regression models}\label{sec-app}

\begin{figure}[t]
\begin{center}
\centerline{\includegraphics[width=3in]{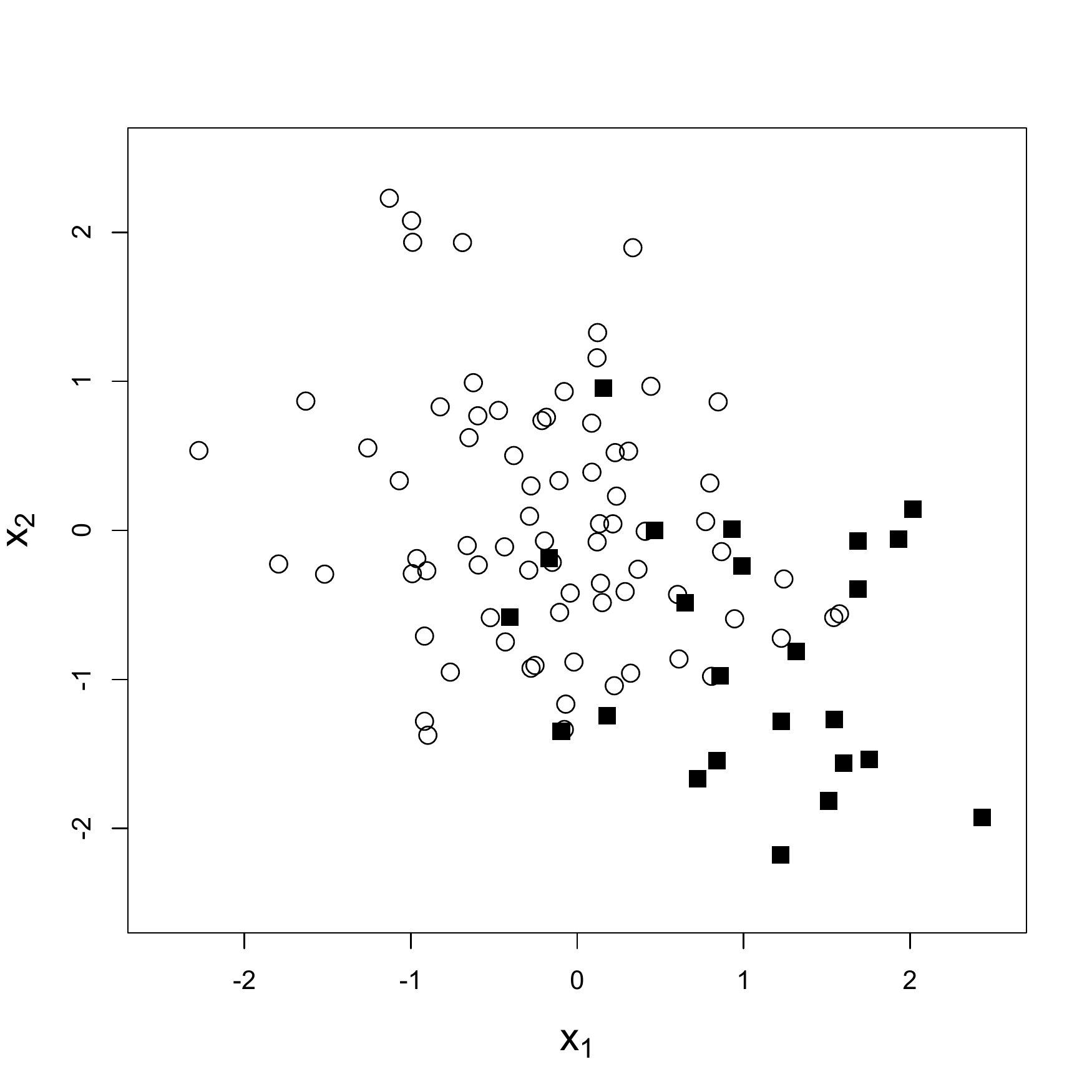}}
\caption{An illustrative binary classification problem with $n=100$ data points 
and two covariates, $x_{1}$ and $x_{2}$, with the two classes represented
by white circles and black squares.} 
\label{illustExample}
\end{center}
\end{figure}

We now look at how Split HMC can be applied to Bayesian logistic
regression models for binary classification problems.  We will
illustrate this method using the simulated data set with $n = 100$ data
points and $p=2$ covariates that is shown in Figure~\ref{illustExample}.

The logistic regression model assigns probabilities to the two possible classes
(denoted by 0 and 1) in case $i$ (for $i=1,\ldots,n$) as follows:
\begin{eqnarray}
P(y_{i}=1\,|\,x_{i}, \alpha,{ \beta}) 
  & = & \frac{\exp(\alpha + x_{i}^{T}{ \beta})}{1+\exp(\alpha+x_{i}^{T}{\beta})}
\end{eqnarray}
Here, $x_{i}$ is the vector of length $p$ with the observed values of the
covariates in case $i$, $\alpha$ is the intercept, and $\beta$ is the 
vector of $p$ regression coefficients. 
We use $\theta$ to denote the vector of all $p+1$ unknown
parameters, $(\alpha,\,{\beta})$. 

Let $P(\theta)$ be the prior
distribution for $\theta$. The posterior distribution of $\theta$
given $x$ and $y$ is proportional to
$P(\theta)\prod_{i=1}^{n}P(y_{i}|x_{i}, \theta)$. The corresponding
potential energy function is
\begin{eqnarray*}
U(\theta) & = & -\log[P(\theta)]\ -\ \sum_{i=1}^{n}\log[P(y_{i}|x_{i}, \theta)]
\end{eqnarray*}
We assume the following (independent) priors for the model parameters:
\begin{eqnarray*}
\alpha & \sim & N(0, \sigma^{2}_{\alpha})\\
\beta_{j} & \sim & N(0, \sigma ^{2}_{\beta}), \qquad j = 1, \ldots, p
\end{eqnarray*}
where $\sigma_{\alpha}$ and $\sigma_{\beta}$ are known constants.

The potential energy function for the above logistic regression model is therefore as follows:
\begin{eqnarray*}
U(\theta) & = & \frac{\alpha^{2}}{2\sigma_{\alpha}^{2}} \ +\
  \sum_{j = 1}^{p}\frac{\beta^{2}_{j}}{2\sigma_{\beta}^{2}}\ -\
  \sum_{i=1}^{n} \Big[ y_i (\alpha + x_{i}^{T} \beta) - \log(1+\exp(\alpha + 
                       x_{i}^{T}\beta)) \Big ] 
\end{eqnarray*}
The partial derivatives of the energy function with respect to $\alpha$
and the $\beta_{j}$ are
\begin{eqnarray*}
\ \ \ \frac{\partial U}{\partial \alpha} \ = \ 
 \frac{\alpha}{\sigma^{2}_{\alpha}} \ -\ 
 \sum_{i=1}^{n} \Big[y_{i} - \frac{\exp(\alpha + 
     x_{i}^{T} \beta)}{1 + \exp(\alpha + x_{i}^{T}\beta)}\Big],\ \ \ 
\frac{\partial U}{\partial \beta_{j}} \ = \ 
 \frac{\beta_{j}}{\sigma^{2}_{\beta}} \ -\ 
 \sum_{i=1}^{n} x_{ij} \Big[y_{i} - \frac{\exp(\alpha + 
     x_{i}^{T} \beta)}{1 + \exp(\alpha + x_{i}^{T}\beta)}\Big]
\end{eqnarray*}

\subsection{Split HMC with a partial analytical solution for a logistic model}

\begin{figure}[t]
\begin{center}
 \centerline{\includegraphics[width=3in]{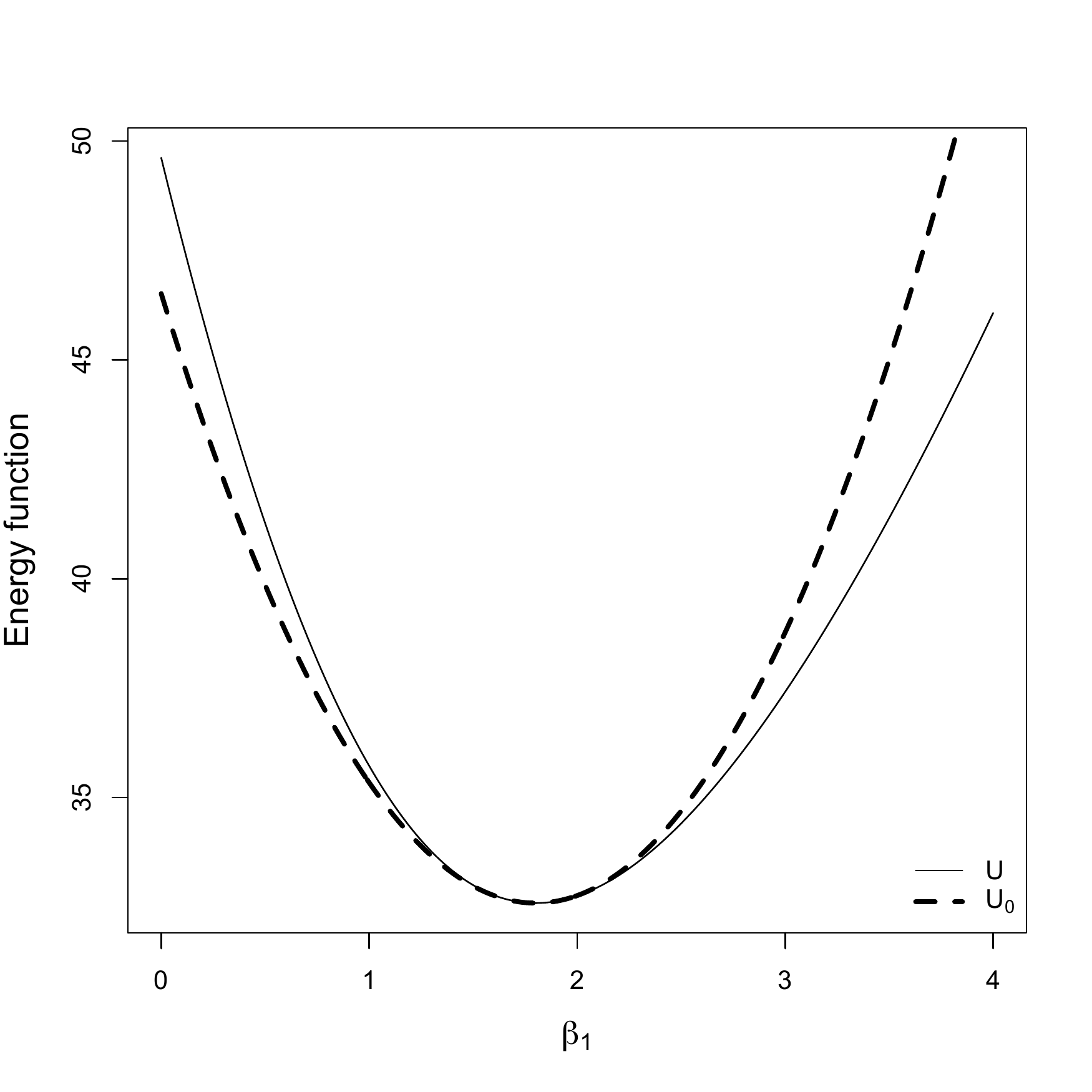}\includegraphics[width=3in]{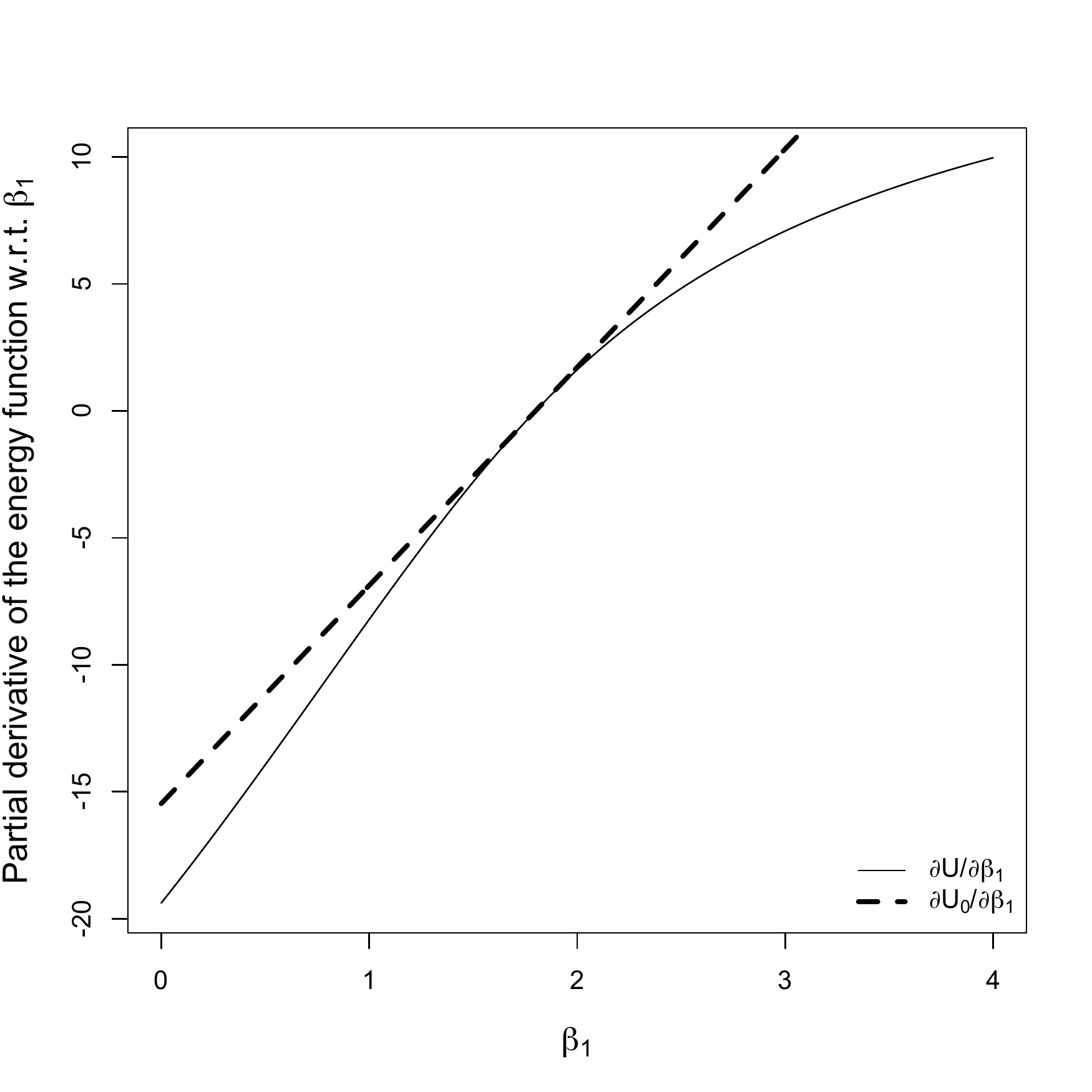}}
\caption{Left plot: The potential energy, $U$, for the logistic regression model (solid curve) and its normal approximation, $U_{0}$ (dashed curve), as $\beta_1$ varies, with other parameters at their
MAP values. Right plot: The partial derivatives of $U$ and $U_{0}$ with respect to $\beta_{1}$.} 
\label{splitNormPlot}
\end{center}
\end{figure}

To apply Algorithm~1 for Split HMC to this problem, we approximate the 
potential energy function $U(\theta)$ for the logistic regression model with 
the potential energy function $U_{0}(\theta)$ of the normal distribution
$N(\hat{\theta}, \mathcal{J}^{-1}(\hat{\theta}))$, where
$\hat{\theta}$ is the MAP estimate of model
parameters. $U_{0}(\theta)$ usually provides a reasonable
approximation to $U(\theta)$, as illustrated in Figure
\ref{splitNormPlot}.  In the plot on the left, the solid curve shows
the value of the potential energy, $U$, as $\beta_1$ varies, with
$\beta_2$ and $\alpha$ fixed to their MAP values, while the dashed
curve shows $U_{0}$ for the approximating normal distribution.  The
right plot of Figure \ref{splitNormPlot} compares the partial
derivatives of $U$ and $U_{0}$ with respect to $\beta_{1}$, showing
that ${\partial U_{0}}/{\partial \beta_{j}}$ provides a reasonable
linear approximation to ${\partial U}/{\partial \beta_{j}}$.

Since there is no error when solving Hamiltonian
dynamics based on $U_{0}(\theta)$, we would expect that the total
discretization error of the steps taken by Algorithm~1 will be
less that for the standard leapfrog method, for a given stepsize, and 
that we will therefore be able to use a larger stepsize --- and hence
need fewer steps for a given trajectory length --- while still
maintaining a good acceptance rate.  The stepsize will still 
be limited to the region of stability imposed by the discretization error 
from $U_{1} = U - U_{0}$, but this limit will tend to be larger than
for the standard leapfrog method.

\subsection{Split HMC with splitting of data for a logistic model}

\begin{figure}[t]
\begin{center}
 \centerline{\includegraphics[width=3in]{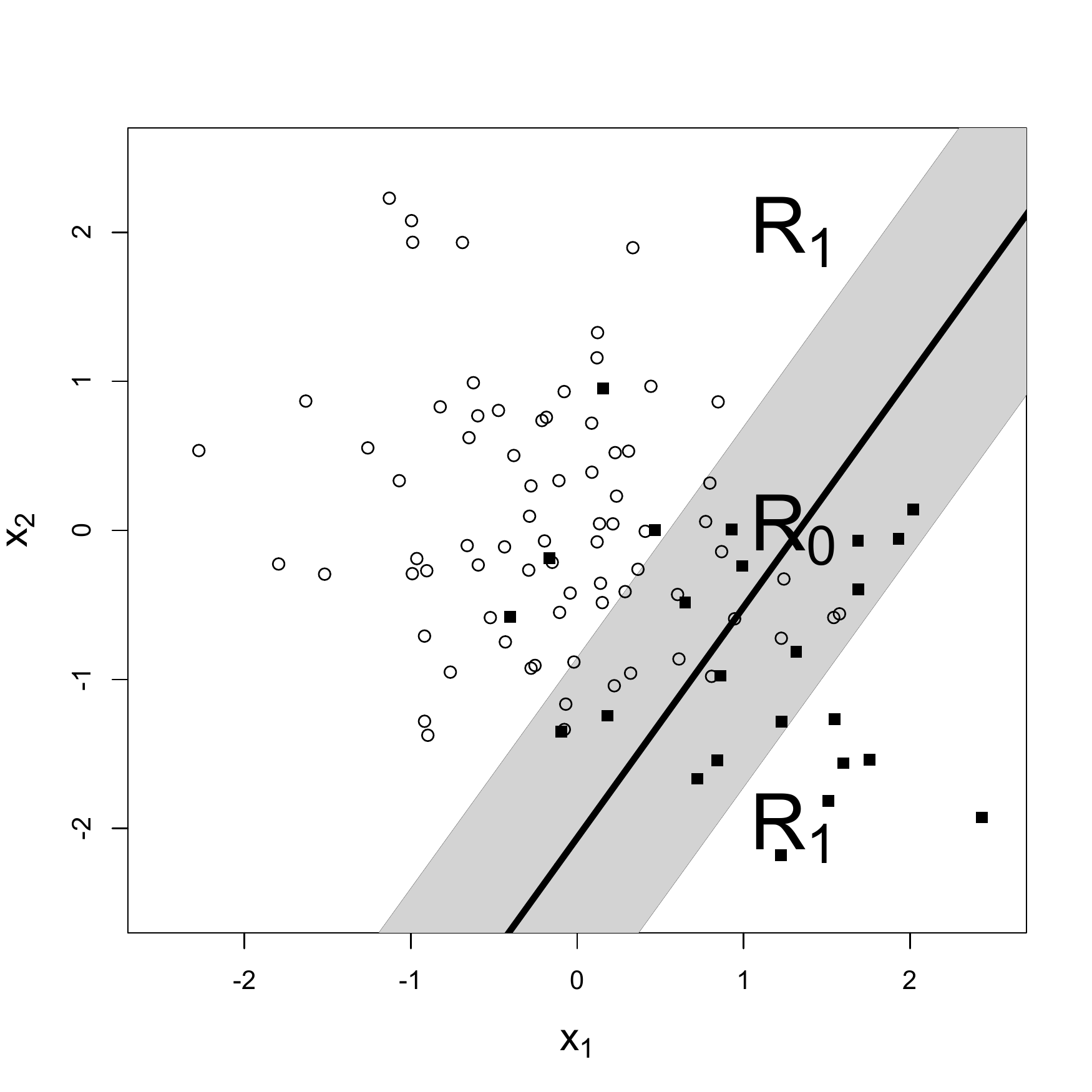} \includegraphics[width=3in]{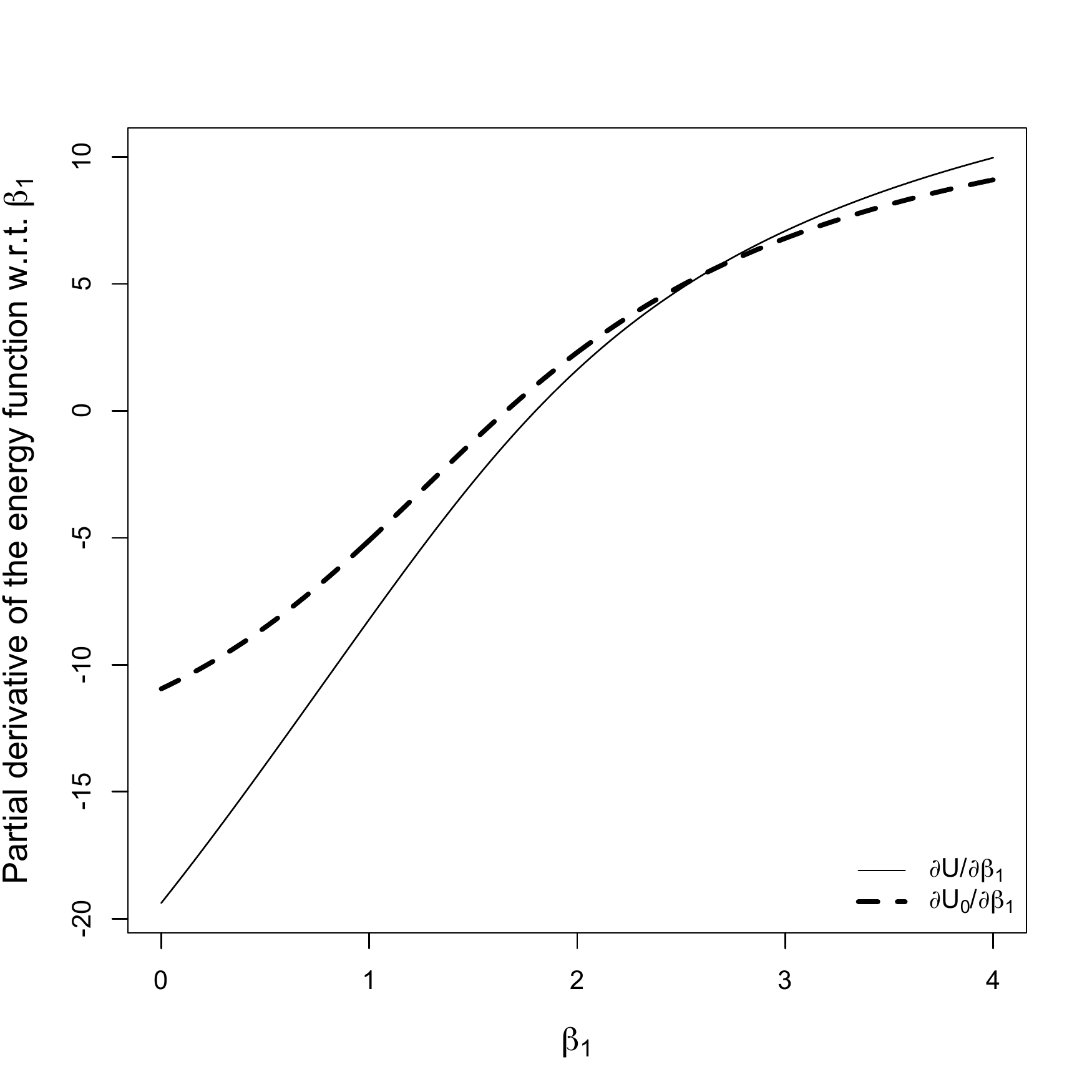}}
\caption{Left plot: A split of the data into two parts based on the
MAP model, represented by the solid line; the energy function $U$ is then
divided into $U_{0}$, based on the data points in $R_{0}$,
and $U_{1}$, based on the data points in $R_{1}$. Right plot:
The partial derivatives of $U$ and $U_{0}$ with respect to
$\beta_{1}$, with other parameters at their MAP values.}
\label{splitDataPlot}
\end{center}
\end{figure}

To apply Algorithm 2 to this logistic regression model, we split the
Hamiltonian by splitting the data into two subsets.  Consider the
illustrative example discussed above. In the left plot of Figure
\ref{splitDataPlot}, the thick line represents the classification
boundary using the MAP estimate, $\hat{\theta}$. For the points that
fall on this boundary line, the estimated probabilities for the two
groups are equal, both being 1/2.  The probabilities of the two
classes become less equal as the distance of the covariates from this
line increases.  We will define $U_0$ using the points within the
region, $R_0$, within some distance of this line, and define $U_1$
using the points in the region, $R_1$, at a greater distance from this line.
Equivalently, $R_0$ contains those points for which the probability
that $y=1$ (based on the  MAP estimates) is closest to 1/2.

The shaded area in Figure \ref{splitDataPlot} shows the region, $R_0$,
containing the 30\% of the observations closest to the MAP line, or 
equivalently the 30\% of observations for which the probability
of class~1 is closest (in either direction) to 1/2.
The unshaded region containing the remaining data points is denoted as
$R_{1}$.  Using these two subsets, we can split the energy function
$U(\theta)$ into two terms: $U_{0}(\theta)$ based on the data points
that fall within $R_{0}$, and $U_{1}$ based on the data points that
fall within $R_{1}$ (see Eq.~\eqref{eq:splitData}). Then, we use
Eq.~\eqref{split3} to split the Hamiltonian dynamics.

Note that $U_{0}$ is not used to approximate the potential energy
function, $U$, for the acceptance test at the end of the trajectory ---
the exact value of $U$ is used for this test, ensuring that the
equilibrium distribution is exactly correct.
Rather, ${\partial U_{0}}/{\partial \beta_{j}}$ is used to approximate
${\partial U}/{\partial \beta_{j}}$, which is the costly
computation when we simulate Hamiltonian dynamics. 

To see that it is appropriate to split the data according to how close
the probability of class~1 is to 1/2, note first that the leapfrog
step of Eq.~(\ref{leapfrog}) will have no error when the derivatives
$\partial U/\partial q_j$ do not depend on $q$ --- that is, when 
the second derivatives of $U$ are zero.  
Recall that for the logistic model,
\begin{eqnarray*}
\frac{\partial U}{\partial \beta_{j}} & = & 
\frac{\beta_{j}}{\sigma^{2}_{\beta}} \ - \
\sum_{i=1}^{n} x_{ij} \left[y_{i} - \frac{\exp(\alpha + x_{i}^{T} \beta)}
{1 + \exp(\alpha + x_{i}^{T}\beta)}\right]
\end{eqnarray*}
from which we get
\begin{eqnarray*}
\frac{\partial^2 U}{\partial \beta_{j}\beta_{j'}} & = & 
\frac{\delta_{jj'}}{\sigma^{2}_{\beta}} \ + \
\sum_{i=1}^{n} x_{ij} x_{ij'} \left[\frac{\exp(\alpha + x_{i}^{T} \beta)}
{1 + \exp(\alpha + x_{i}^{T}\beta)}
- \left(\frac{\exp(\alpha + x_{i}^{T} \beta)}
{1 + \exp(\alpha + x_{i}^{T}\beta)}\right)^2\right]\\[4pt]
& = & 
\frac{\delta_{jj'}}{\sigma^{2}_{\beta}} \ + \
\sum_{i=1}^{n} x_{ij} x_{ij'} P(y_i=1|x_i,\alpha,\beta)
                      [1 - P(y_i=1|x_i,\alpha,\beta)]
\end{eqnarray*}
The product $P(y_i=1|x_i,\alpha,\beta)[1 - P(y_i=1|x_i,\alpha,\beta)]$
is symmetrical around its maximum where 
$P(y_i=1|x_i,\alpha,\beta)$ is 1/2, justifying 
our criterion for selecting points in $R_0$.
The right plot of Figure \ref{splitDataPlot} shows
the approximation of ${\partial U}/{\partial \beta_{1}}$ by ${\partial
U_{0}}/{\partial \beta_{1}}$ with $\beta_2$ and $\alpha$ fixed to their MAP values.

\subsection{Experiments}\label{sec-exp}

In this section, we use simulated and real data to compare our
proposed methods to standard HMC. For each problem, we set the
number of leapfrog steps to $L=20$ for standard HMC, and find
$\epsilon$ such that the acceptance probability (AP) is close to 0.65
\citep{neal10}.  We set $L$ and $\epsilon$ for the Split HMC methods
such that the trajectory length, $\epsilon L$, remains the same, but
with a larger stepsize and hence a smaller number of steps.  Note that
this trajectory length is not necessarily optimal for these problems,
but this should not affect our comparisons, in which the length is
kept fix.

We try to choose $\epsilon$ for the Split HMC methods such that the
acceptance probability is equal to that of standard HMC.  However,
increasing the stepsize beyond a certain point leads to instability of
trajectories, in which the error of the Hamiltonian grows rapidly with
$L$ \citep{neal10}, so that proposals are rejected with very high
probability.  This sometimes limits the stepsize of Split HMC to
values at which the acceptance probability is greater than the 0.65
aimed at for standard HMC.  Additionally, to avoid near periodic
Hamiltonian dynamics \citep{neal10}, we randomly vary the stepsize
over a small range.  Specifically, at each iteration of MCMC, we
sample the stepsize from the Uniform$(0.8 \epsilon, \epsilon)$
distribution, where $\epsilon$ is the reported stepsize for each
experiment.

To measure the efficiency of each sampling method, we use the autocorrelation time (ACT) by dividing the $N$ posterior samples into batches of size $B$, and estimating ACT as follows \citep{neal93, geyer92}:
\begin{eqnarray*}
\tau & = & B \frac{S^{2}_{b}}{S^{2}}
\end{eqnarray*}
Here, $S^{2}$ is the sample variance and $S^{2}_{b}$ is the sample variance of batch means. Following \cite{ACT}, we divide the posterior samples into $N^{1/3}$ batches of size $B = N^{2/3}$. Throughout this section, we set the number of Markov chain Monte Carlo (MCMC) iterations for simulating posterior samples to $N=50000$.

The autocorrelation time can be roughly interpreted as the number of
MCMC transitions required to produce samples that can be considered as
independent. For the logistic regression problems discussed in this
section, we could find the autocorrelation time separately for each parameter and summarize the autocorrelation times using their maximum value (i.e., for the slowest moving parameter) to compare different methods. However, since one common goal is to use logistic regression models for prediction, we look at the autocorrelation time for the log likelihood, $\sum_{i=1}^{n} \log[P(y_{i} | x_{i}, \theta)]$ using the posterior samples of $\theta$. We also look at the autocorrelation time for $\sum_{j} (\beta_j)^{2}$ (denoting it $\tau_{\beta}$), since this may be more relevant when the goal is interpretation of parameter estimates.

We adjust $\tau$ (and similarly $\tau_{\beta}$) to account for the varying computation time
needed by the different methods in two ways.  One is to compare
different methods using $\tau \times s$, where $s$ is the CPU
time per iteration, using an implementation written in R.  This
measures the CPU time required to produce samples that can be regarded
as independent samples.  We also compare in terms of $\tau
\times g$, where $g$ is the number of gradient computations on the
number of cases in the full data set required for each trajectory
simulated by HMC.  This will be equal to the number of leapfrog steps,
$L$, for standard HMC or Split HMC using a normal approximation.  When
using data splitting with a fraction $f$ of data in $R_0$ and $M$
inner leapfrog steps, $g$ will be $(fM+(1-f))\times L$. In general,
we expect that computation time will be dominated by the gradient
computations counted by $g$, so that $\tau\times g$ will
provide a measure of performance independent of any particular
implementation.  In our experiments, $s$ was close to being
proportional to $g$, except for slightly larger than expected times
for Split HMC with data splitting.

\subsubsection{Simulated data}

\begin{table}[t]
\begin{center}
{\scriptsize{
\begin{tabular}{l||c|c|c|}
& HMC & \multicolumn{2}{c|}{Split HMC}\\
& & Normal Appr. & Data Splitting \\
\hline \hline 
$L$ & 20 & 10 & 3\\
$g$ & 20 & 10 & 12.6\\
$s$ & 0.187 & 0.087 & 0.096 \\
$AP$ & 0.69 & 0.74 & 0.74\\
$\tau$ & 4.6 & 3.2& 3.0\\
\rowcolor[gray]{.8} $\tau \times g $ &  92 & {\bf{32}} & 38 \\
\rowcolor[gray]{.8} $\tau \times s $ & 0.864 & {\bf{0.284}} & 0.287 \\
$\tau_{\beta}$ & 11.7 & 13.5 & 7.3 \\
\rowcolor[gray]{.8} $\tau_{\beta} \times g $ &  234 & 135 & {\bf{92}} \\
\rowcolor[gray]{.8} $\tau_{\beta} \times s $ & 2.189 & 1.180 & {\bf{0.703}} \\
\hline
\end{tabular}
}}
\end{center}
\caption{Split HMC (with normal approximation and data
splitting) compared to standard HMC using simulated data,
on a data set with $n=10000$ observations and $p=100$ covariates.}  
\label{sim1Results} \end{table}%
  
We first tested the methods on a simulated data set with 100 covariates 
and 10000 observations.  The covariates were
sampled as $x_{ij} \sim N(0, \sigma^{2}_{j})$, for $i=1,
\ldots, 10000$ and $j=1, \ldots, 100$, with $\sigma_j$ set to 5 for the
first five variables, to 1 for the next five variables, and to 0.2 for
the remaining 90 variables. We sampled true parameter values, $\alpha$
and $\beta_j$, independently from $N(0,1)$ distributions.  Finally,
we sampled the class labels according to the model, as $y_i 
\sim \mbox{Bernoulli}(p_{i})$ with 
$\textrm{logit}(p_{i}) \, = \, \alpha + x_{i}^{T}\beta$.

For the Bayesian logistic regression model, we assumed normal priors
with mean zero and standard deviation 5 for $\alpha$ and $\beta_{j}$,
where $j=1, \ldots, 100$. We ran standard HMC, Split HMC with normal
approximation, and Split HMC with data splitting for $N=50000$
iterations. For the standard HMC, we set $L = 20$ and
$\epsilon=0.015$, so the trajectory length was $20 \times 0.015 =
0.3$. For Split HMC with normal approximation and Split HMC with data
splitting, we reduce the number of leapfrog steps to 10 and 3
respectively, while increasing the stepsizes so that the trajectory
length remained $0.3$. For the data splitting method, we use 40\% of
the data points for $U_{0}$ and set $M=9$, which makes $g$ equal
$4.2L$.  For this example, we set $L=3$ so $g=12.6$, which is smaller
than $g=L=20$ used for the standard HMC algorithm

Table \ref{sim1Results} shows the results for the three methods. The
CPU times (in seconds) per iteration, $s$, and $\tau \times s$
for the Split HMC methods are substantially lower than for standard
HMC. The comparison
is similar looking at $\tau \times g$. Based on $\tau_{\beta} \times s$ and $\tau_{\beta} \times g$, however, the improvement in efficiency is more substantial for the data splitting method compared to the normal approximation method mainly because of the difference in their corresponding values of $\tau_{\beta}$.

\subsubsection{Results on real data sets}

\begin{table}[p]
\begin{center}
{\scriptsize{
\begin{tabular}{l l || l|c|c|c|}
& & HMC & \multicolumn{2}{c|}{Split HMC}\\
& & & Normal Appr. & Data Splitting \\
\hline \hline 
StatLog \qquad & $L$ & 20 & 14 & 3 \\
$n=4435$, $p=37$ \qquad \qquad & $g$ & 20 & 14 & 13.8 \\
& $s$ & 0.033 & 0.026 & 0.023 \\
& $AP$ & 0.69 & 0.74 & 0.85\\
& $\tau$ & 5.6 &  6.0 & 4.0\\
\rowcolor[gray]{.8} & $\tau \times g $ & 112  & 84  & {\bf{55}}\\
\rowcolor[gray]{.8} & $\tau \times s $ & 0.190  & 0.144  & {\bf{0.095}}\\
& $\tau_{\beta}$ & 5.6 & 4.7 & 3.8 \\
\rowcolor[gray]{.8} & $\tau_{\beta} \times g $ &  112 & 66 & {\bf{52}} \\
\rowcolor[gray]{.8} & $\tau_{\beta} \times s $ & 0.191 & 0.122 & {\bf{0.090}} \\
\hline
CTG \qquad & $L$ & 20 & 13 & 2\\
$n=2126$, $p=21$ & $g$ & 20 & 13 & 9.8 \\
& $s$ & 0.011 & 0.008 & 0.005 \\
& $AP$ & 0.69 & 0.77 & 0.81\\
& $\tau$ & 6.2 & 7.0 & 5.0\\
\rowcolor[gray]{.8} & $\tau \times g $ & 124  & 91 & {\bf{47}}\\
\rowcolor[gray]{.8} & $\tau \times s $ & 0.069  & 0.055  & {\bf{0.028}}\\
& $\tau_{\beta}$ & 24.4 & 19.6 & 11.5 \\
\rowcolor[gray]{.8} & $\tau_{\beta} \times g $ &  488 & 255 & {\bf{113}} \\
\rowcolor[gray]{.8} & $\tau_{\beta} \times s $ & 0.271 & 0.154 & {\bf{0.064}} \\
\hline
Chess \qquad & $L$ & 20 & 9 & 2 \\
$n=3196$, $p=36$ & $g$ & 20 & 13 & 11.8 \\
& $s$ & 0.022 & 0.011 & 0.013 \\
& $AP$ & 0.62 & 0.73 & 0.62\\
& $\tau$ & 10.7 & 12.8 & 12.1\\
\rowcolor[gray]{.8} & $\tau \times g $ &  214 & {\bf{115}}  & 143 \\
\rowcolor[gray]{.8} & $\tau \times s $ & 0.234  & {\bf{0.144}}  & 0.161 \\
& $\tau_{\beta}$ & 23.4 & 18.9 & 19.0 \\
\rowcolor[gray]{.8} & $\tau_{\beta} \times g $ &  468 & 246 & {\bf{224}} \\
\rowcolor[gray]{.8} & $\tau_{\beta} \times s $ & 0.511 & {\bf{0.212}} & 0.252 \\
\hline
\end{tabular}
}}
\end{center}
\caption{HMC and Split HMC (normal approximation and data splitting) on 
three real data sets.}
\label{realData}
\end{table}%

In this section, we evaluate our proposed method using three real
binary classification problems. The data for these three problems are
available from the UCI Machine Learning Repository
(\url{http://archive.ics.uci.edu/ml/index.html}). For all data sets, we standardized the numerical variables to have mean zero and standard deviation 1. Further, we assumed normal priors with mean zero and standard deviation 5 for the regression parameters. We used the setup described at the beginning of Section~\ref{sec-exp}, running each
Markov chain for $N=50000$ iterations. Table \ref{realData}
summarizes the results using the three sampling methods.

The first problem, StatLog, involves using multi-spectral values of
pixels in a satellite image in order to classify the associated area
into soil or cotton crop. (In the original data, different types of
soil are identified.)  The sample size for this data set is $n=4435$,
and the number of features is $p=37$. For the standard HMC, we set
$L=20$ and $\epsilon= 0.08$. For the two Split HMC methods with
normal approximation and data splitting, we reduce $L$ to 14 and 3
respectively while increasing $\epsilon$ so $\epsilon \times L$
remains the same as that of the standard HMC. For the data splitting methods, we use 40\% of data points for $U_{0}$ and set $M = 10$. As seen in the table,
the Split HMC methods improve efficiency with the data splitting method performing substantially
better than the normal approximation method.

The second problem, CTG, involves analyzing 2126 fetal cardiotocograms
along with their respective diagnostic features \citep{compos00}. The
objective is to determine whether the fetal state class is
``pathologic'' or not. The data include $2126$ observations and $21$
features. For the standard HMC, we set $L=20$ and $\epsilon=
0.08$. We reduced the number of leapfrog steps to 13 and 2 for Split
HMC with normal approximation and data splitting respectively. For the latter, we use 30\% of data points for $U_{0}$ and set $M=14$. Both splitting methods improved performance significantly. As before, the data splitting method outperforms the normal approximation method. 

The objective of the last problem, Chess, is to predict chess endgame
outcomes --- either ``white can win'' or ``white cannot win''. This
data set includes $n=3196$ instances, where each instance is a
board-descriptions for the chess endgame. There are $p=36$ attributes
describing the board. For the standard HMC, we set $L=20$ and
$\epsilon= 0.09$. For the two Split HMC methods with normal
approximation and data splitting, we reduced $L$ to 9 and 2
respectively. For the data splitting method, we use 35\% of the data points for $U_{0}$ and set $M=15$. Using the Split HMC methods, the computational efficiency is improved substantially compared to standard HMC. This time however, the normal approximation approach performs better than the data splitting method in terms of $\tau \times g$, $\tau \times s$, and $\tau_{\beta} \times s$, while the latter performs better in terms of $\tau_{\beta} \times g$.

\section{Discussion}

We have proposed two new methods for improving the efficiency of
HMC, both based on splitting the Hamiltonian in a way that allows
much of the movement around the state space to be performed at
low computational cost. 

While we demonstrated our methods on binary logistic regression models, 
they can be extended to multinomial logistic (MNL) models for multiple 
classes. For MNL models, the regression
parameters for $p$ covariates and $J$ classes form a matrix of $(p+1)$
rows and $J$ columns, which we can vectorize so that the model
parameters, $\theta$, become a vector of $(p+1)\times J$
elements. For Split HMC with normal approximation, we can define
$U_{0}(\theta)$ using an approximate multivariate normal
$N(\hat{\theta}, \mathcal{J}^{-1}(\hat{\theta}))$ as before. For Split
HMC with data splitting, we can still construct $U_{0}(\theta)$ using
a small subset of data, based on the class probabilities for each
data item found using the MAP estimates for the parameters
(the best way of doing this is a subject for future research).
The data splitting method could be further extended to any model
for which it is feasible to find a MAP estimate, and then divide the data 
into two parts based on ``residuals'' of some form.

Another area for future research is to look for tractable
approximations to the posterior distribution other than normal
distributions.  One could also investigate other methods for splitting
the Hamiltonian dynamics by splitting the data --- for example, 
fitting a support vector machine (SVM) to binary classification data, and
using the support vectors for constructing $U_{0}$. 

While the results on simulated data and real problems
presented in this paper have demonstrated the advantages of splitting
the Hamiltonian dynamics in terms of improving the sampling
efficiency, our proposed methods do require preliminary analysis of
data, mainly, finding the MAP estimate. The performance of our
approach obviously depends on how well the corresponding normal
distribution based on MAP estimates approximates the posterior
distribution, or how well a small subset of data found using this MAP
estimate captures the overall patterns in the whole data set.  This
preliminary analysis involves some computational overhead, but the
computational cost associated with finding the MAP estimate is often
negligible compared to the potential improvement in sampling
efficiency for the full Bayesian model.  For the data splitting
method, one could also consider splitting based on the class probabilities
produced by a model that is simpler than the one being fitted using HMC.

Another approach to improving HMC has recently been proposed by
\cite{girolami11}. Their method, Riemannian Manifold HMC (RMHMC), can
also substantially improve performance.  RMHMC utilizes the geometric
properties of the parameter space to explore the best direction,
typically at higher computational cost, to produce distant proposals
with high probability of acceptance. In contrast, our method attempts
to find a simple approximation to the Hamiltonian to reduce the
computational time required for reaching distant states. It is
possible that these approaches could be combined, to produce a method
that performs better than either method alone.  The recent proposals
of \cite{hoffman-gelman} for automatic tuning of HMC could also be combined
with our Split HMC methods.

\end{document}